\documentclass{andp2012}
\usepackage[english]{babel}
\usepackage{graphicx}
\usepackage{amsmath}
\usepackage{amssymb}
\usepackage{bm}
\usepackage{hyperref}

\usepackage{color}

\makeatletter
\providecommand\@enum@widestlabel{7}
\makeatother

\title{Pfaffian formula for fermion parity fluctuations in a superconductor and application to Majorana fusion detection}
\author[A. Grabsch]{A. Grabsch}
\author[Y. Cheipesh]{Y. Cheipesh\inst{}}
\author[C.W.J. Beenakker]{C.W.J. Beenakker\inst{}}
\address[]{Instituut-Lorentz, Universiteit Leiden, P.O. Box 9506, 2300 RA Leiden, The Netherlands}
\shortauthors{Grabsch, Cheipesh \& Beenakker}
\begin{abstract}
Kitaev's Pfaffian formula equates the ground-state fermion parity of a closed system to the sign of the Pfaffian of the Hamiltonian in the Majorana basis. Using Klich's theory of counting statistics for paired fermions we generalize the Pfaffian formula to account for quantum fluctuations in the fermion parity of an open subsystem. A statistical description in the framework of random-matrix theory is used to answer the question when a vanishing fermion parity in a superconductor fusion experiment becomes a distinctive signature of an isolated Majorana zero-mode.
\end{abstract}
\shortabstract
\begin{document}
\maketitle

\section{Introduction}
\label{intro}

The pairing interaction in a superconductor favors a ground state with an \textit{even} number of electrons, but when both time-reversal and spin-rotation symmetry are broken the ground state may have \textit{odd} parity --- for example when a magnetic impurity binds an unpaired electron \cite{Bal06}. While the connection between fermion-parity switches and level crossings was noticed already in 1970 by Sakurai \cite{Sak70}, these only became a topic of intense research activity after Kitaev \cite{Kit01} made the connection with topological phase transitions and Majorana fermions: The absence of level repulsion at a fermion-parity switch indicates a change in a topological quantum number, which Kitaev identified as the sign of the Pfaffian of the Hamiltonian in the basis of Majorana fermions.

\begin{figure}[tb]
\centerline{\includegraphics[width=0.8\linewidth]{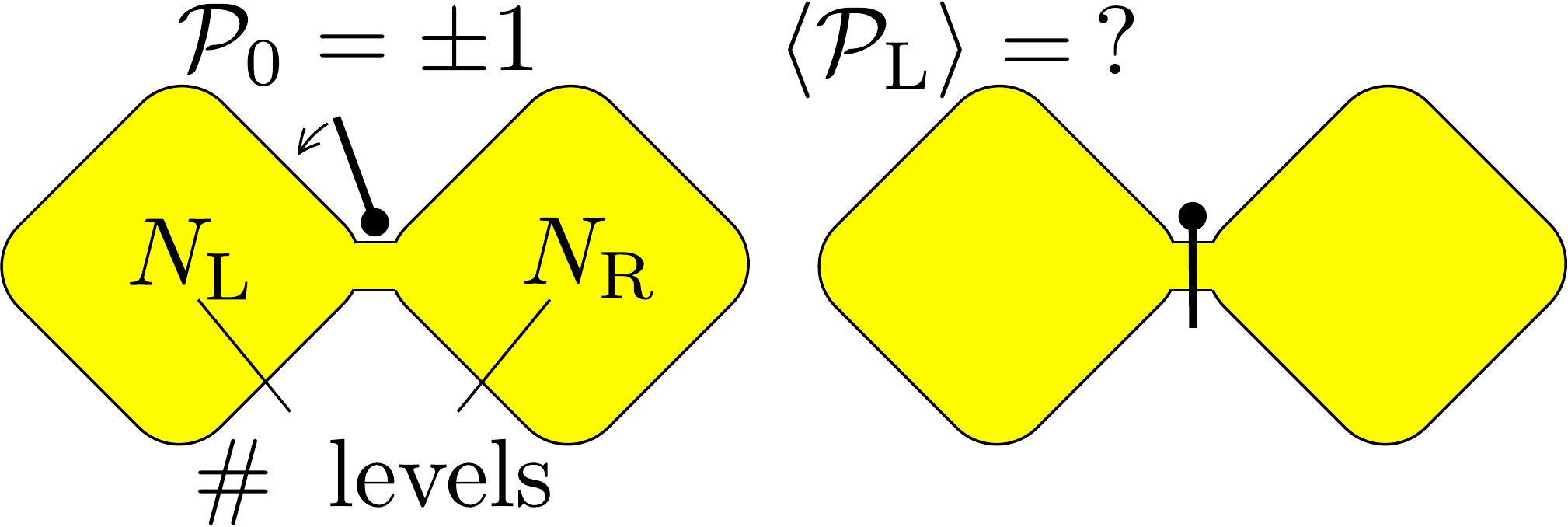}}
\caption{The left panel shows two superconducting regions (quantum dots) connected (fused) by a point contact. The entire system is in a state of definite fermion parity ${\cal P}_{0}$, even ($+1$) or odd ($-1$). The parity ${\cal P}_{\rm L}$ of the occupation number of the $N_{\rm L}$ electronic levels in one single quantum dot has quantum fluctuations. The expectation value $\langle{\cal P}_{\rm L}\rangle\in[-1,1]$ may be obtained by rapidly closing the point contact and decoupling the quantum dots (right panel), followed by a measurement of the fermion parity of a single dot. The effective number of levels $N_{\rm dot}\simeq \hbar/\delta_0 \tau_{c}$ in each quantum dot that contributes to the fermion parity fluctuations is determined by the single-particle level spacing $\delta_0$ and the time scale $\tau_{c}$ on which the interdot coupling is broken \cite{Cla17}. We address the question when a vanishing fermion parity $\langle{\cal P}_{\rm L}\rangle\approx 0$ in such a fusion experiment is a signature of isolated Majorana zero-modes.
}
\label{fig_diagram}
\end{figure}

An open subsystem need not be in a state of definite fermion parity ${\cal P}=\pm 1$, the fermion parity expectation value $\langle {\cal P}\rangle$ may take on any value in the interval $[-1,1]$. Here we generalize Kitaev's Pfaffian formula so that it can describe both closed and open systems. This generalization has a computational as well as a conceptual merit. Computationally, it reduces the complexity of a calculation of $\langle {\cal P}\rangle$ for $N$ levels from order $2^N$, when all possible occupation numbers are enumerated, down to order $N^3$. Conceptually, it allows us to make contact with the random-matrix theory of topological superconductivity \cite{Alt97,Bee15}, and identify the origin of a statistical peak at $\langle{\cal P}\rangle=0$ discovered recently in computer simulations \cite{Cla17}. These findings have implications for proposed experiments \cite{Aas16} to search for signatures of isolated Majorana zero-modes in the fermion parity of two superconductors that have first been fused and then decoupled (see Fig.\ \ref{fig_diagram}).

The outline of the paper is as follows. In the next section we derive the Pfaffian formula for the average subsystem fermion parity. This generalization of Kitaev's formula \cite{Kit01} can be seen either as an application of the Wick theorem for Majorana operators \cite{Per77,Per84,Bra05} (cf.\ a similar application in Ref.\ \citen{Bau18}), or as an application of Klich's theory of counting statistics for paired fermions \cite{Kli14}. In Sec.\ \ref{sec_fusionrule} we use the fermion parity formula to establish the connection between vanishing average fermion parity and the presence of isolated Majorana zero-modes in the decoupled quantum dot. We continue in Sec.\ \ref{sec_RMT} with a statistical description of the double quantum dot geometry of Fig.\ \ref{fig_diagram}, by identifying the random-matrix ensemble in symmetry class DIII that describes the fermion parity fluctuations. We contrast the case of strongly coupled quantum dots in Sec.\ \ref{sec_PLdistribution} with the case of weak coupling in Sec.\ \ref{sec_coupling}. In Sec.\ \ref{sec_isolated} we show how weak coupling by a single-mode quantum point contact can distinguish quantum dots with or without isolated Majorana zero-modes. In the concluding Sec.\ \ref{sec_discuss} we discuss the implications of our analysis for the detection of Majorana zero-modes by means of a fusion experiment.

\section{Pfaffian fermion-parity formula}
\label{sec_Pfaffianformula}

\subsection{Kitaev's formula for an isolated system}
\label{sec_Kitaev}

To set the stage we recall some basic facts \cite{Bud13} needed to present Kitaev's formula \cite{Kit01} for the ground-state fermion parity of an isolated superconductor. 

At the mean-field level the Hamiltonian of a superconductor is a Hermitian quadratic form in the fermion creation and annihilation operators $a^\dagger$, $a$,
\begin{align}
H={}&\textstyle{\sum_{n,m=1}^N}V_{nm}\left(a^\dagger_n a^{\vphantom{\dagger}}_m-\tfrac{1}{2}\delta_{nm}\right)\nonumber\\
&+\tfrac{1}{2}\textstyle{\sum_{n,m=1}^N}\left(\Delta_{nm}a_na_m+\Delta_{nm}^\ast a_m^\dagger a_n^\dagger\right).\label{Haa}
\end{align}
The indices $n,m$ label spin and orbital degrees of freedom of $N$ fermionic modes. The $N\times N$ Hermitian matrix $V$ represents the kinetic and potential energy and the antisymmetric matrix $\Delta$ is the pair potential. 

More compactly, Eq.\ (\ref{Haa}) can be written in the matrix form
\begin{subequations}
\label{HBdef}
\begin{align}
&H=\tfrac{1}{2}\sum_{n,m=1}^N\Psi_n^\dagger\cdot{\cal B}_{nm}\cdot\Psi_m,\label{HBdefa}\\
&\Psi_n=\begin{pmatrix}
a_n \\
a_n^\dagger
\end{pmatrix},\;\;{\cal B}_{nm}=\begin{pmatrix}
V_{nm}&-\Delta^\ast_{nm}\\
\Delta_{nm}&-V_{nm}^\ast
\end{pmatrix}.\label{HBdefb}
\end{align}
\end{subequations}
The $2N\times 2N$ Hermitian matrix ${\cal B}$ is called the Bogoliubov-De Gennes (BdG) Hamiltonian \cite{DeG66}. Its eigenvalues come in pairs $\pm E_{1}, \pm E_{2},\ldots \pm E_{N}$ of opposite sign, with the positive entries equal to the single-particle excitation energies of the many-particle Hamiltonian $H$.

The unitary transformation 
\begin{equation}
{\cal B}_{nm}\mapsto U{\cal B}_{nm}U^\dagger\equiv {\cal A}_{nm},\;\;\text{with}\;\;U=\frac{1}{\sqrt 2}\begin{pmatrix}
1&1\\
-i&i
\end{pmatrix},\label{Udef}
\end{equation}
maps ${\cal B}$ onto the $2N\times 2N$ imaginary antisymmetric matrix ${\cal A}$ with elements
\begin{equation}
{\cal A}_{nm}=\begin{pmatrix}
i\,{\rm Im}\,(V_{nm}+\Delta_{nm})&i\,{\rm Re}\,(\Delta_{nm}+V_{nm})\\
i\,{\rm Re}\,(\Delta_{nm}-V_{nm})&i\,{\rm Im}\,(V_{nm}-\Delta_{nm})
\end{pmatrix}=-{\cal A}_{mn}^{\rm T}.\label{Aresult}
\end{equation}
The superscript T denotes the transpose. An antisymmetric matrix is also referred to as ``skew-symmetric''. 

The transformed state
\begin{equation}
\gamma=(\gamma_1,\gamma_2,\ldots\gamma_{2N}),\;\;\text{with}\;\;\begin{pmatrix}
\gamma_{2n-1}\\
\gamma_{2n}
\end{pmatrix}=U\begin{pmatrix}
a_{n}\\
a^\dagger_n
\end{pmatrix},\label{gammaUdef}
\end{equation}
contains $2N$ Hermitian operators $\gamma_n=\gamma_n^\dagger$, with anticommutator
\begin{equation}
\gamma_n\gamma_m+\gamma_m\gamma_n=\delta_{nm},\;\;\gamma_n^2=1/2.\label{Clifford}
\end{equation}
This is the Clifford algebra of Majorana operators.

The global fermion parity operator 
\begin{equation}
{\cal P}=(-1)^{\sum_{n=1}^N a^\dagger_n a^{\vphantom{\dagger}}_n}=(-2i)^N\gamma_1\gamma_2\cdots\gamma_{2N}\label{Pglobal}
\end{equation}
commutes with $H$, so energy eigenstates have a definite fermion parity $\pm 1$. Kitaev's formula \cite{Kit01} equates the fermion parity ${\cal P}_{0}$ of the ground state to the Pfaffian\footnote{Wikipedia has a helpful collection of Pfaffian formulas.} (Pf) of the Hamiltonian in the Majorana basis,
\begin{equation}
{\cal P}_{0}=\text{sign}\,\text{Pf}\,(-i{\cal A}),\;\;\text{for}\;\;H=\tfrac{1}{2}\gamma\cdot{\cal A}\cdot\gamma.\label{P0Adef}
\end{equation}


\subsection{Pfaffian formula for a subsystem}
\label{sec_subsystem}

Our objective is to calculate the ground-state expectation value of the fermion parity ${\cal P}_{\rm L}$ of an open subsystem, say the left quantum dot with $N_{\text{L}}$ fermionic modes in Fig.\ \ref{fig_diagram} .

A direct way to proceed, used for example in Ref.\ \citen{Cla17}, is to calculate the many-particle ground state $|\Psi_0\rangle$ in the basis of occupation numbers and evaluate
\begin{equation}
\langle{\cal P}_{\rm L}\rangle=\langle\Psi_0|(-1)^{\sum_{n=1}^{N_{\rm L}}a^\dagger_n a^{\vphantom{\dagger}}_n}|\Psi_0\rangle.\label{PLdef}
\end{equation}
Since the Fock space of occupation numbers has dimension $2^{N}$, this direct approach scales exponentially with system size and is therefore prohibitively expensive for large systems.

Klich \cite{Kli14} has developed an efficient method, with a polynomial scaling in $N$, to calculate squares of expectation values of operators $\exp(i\chi\sum_{n}a^\dagger_n a^{\vphantom{\dagger}}_n)$. This gives $\langle{\cal P}_{\rm L}\rangle^2$ if one sets $\chi=\pi$ and restricts the sum to indices $n$ in L. In App.\ \ref{App_Klich} we show how the Klich method can be adapted to give also the sign of $\langle{\cal P}\rangle_{\rm L}$. That calculation is technically rather involved, but the final result can be easily understood as follows.

We make the flat-band transformation ${\cal A}\mapsto\bar{\cal A}$, which consists in replacing each of the $2N$ eigenvalues $\pm E_n$ of ${\cal A}$ by their sign. (We assume that no eigenvalue is identically zero, meaning that we are not precisely at a fermion-parity switch.) Since no eigenvalue crosses zero when it is replaced by its sign, the flat-band transformation leaves the sign of the Pfaffian (\ref{P0Adef}) invariant. And because the Pfaffian of $-i\bar{\cal A}$ can only equal $\pm 1$ we no longer need to take the sign in Eq.\ (\ref{P0Adef}), hence the global fermion parity is \begin{equation}
{\cal P}_0=\text{Pf}\,(-i\bar{\cal A}).\label{Kitaevnosign}
\end{equation}

At this point one may surmise that the desired subsystem generalization of Eq.\ (\ref{P0Adef}) simply amounts to taking the Pfaffian of the $2N_{\rm L}\times 2N_{\rm L}$ submatrix $[\bar{\cal A}]_{\rm LL}$ restricted to the subspace of modes in the left quantum dot,
\begin{equation}
\langle {\cal P}_{\rm L}\rangle=\text{Pf}\,[-i\bar{\cal A}]_{\rm LL}.\label{generalizedPL}
\end{equation}
This is indeed the correct expression, as one can see by application of the Wick theorem for Majorana operators \cite{Per77,Per84,Bra05},
\begin{align}
\langle \gamma_1\gamma_2\cdots\gamma_{2s}\rangle=\underset{1\leq k<l\leq 2s}{\rm Pf}\,\langle\gamma_k\gamma_l\rangle.\label{Wick}
\end{align}
Substitution of ${\cal P}_{\rm L}=(-2i)^ {N_{\rm L}}\gamma_1\gamma_2\cdots\gamma_{2N_{\rm L}}$ on the left-hand-side and $-2i\langle\gamma_k\gamma_l\rangle=-i\bar{\cal A}_{kl}$ on the right-hand-side results in Eq.\ (\ref{generalizedPL}). This is how an equivalent formula was derived recently for a different problem \cite{Bau18}.

Eq.\ (\ref{generalizedPL}) is computationally efficient because the Pfaffian of an $N\times N$ matrix can be calculated in a time that scales polynomially with $N$ \cite{Rub11,Wim12}: It has the same ${\cal O}(N^3)$ complexity as the eigenvalue decomposition one needs for the flat-band transformation ${\cal A}\mapsto\bar{\cal A}$. Note that the flat-band transformation needs to be performed \textit{before} the subblock restriction $\bar{\cal A}\mapsto[\bar{\cal A}]_{\rm{LL}}$ --- the two operations do not commute.

\section{Connection with the Majorana fusion rule}
\label{sec_fusionrule}

As a fundamental application of Eq.\ (\ref{generalizedPL}), consider the case that each quantum dot in Fig.\ \ref{fig_diagram} has a single electronic mode ($N_{\rm L}=N_{\rm  R}=1$), each consisting of two Majorana modes with inter-dot coupling matrix $\Gamma$ but vanishing intra-dot coupling --- so these become fully isolated zero-modes when the quantum dots are decoupled. The Hamiltonian in the Majorana basis is
\begin{equation}
{\cal A}=\begin{pmatrix}
0&i\Gamma\\
-i\Gamma^{\rm T}&0
\end{pmatrix}.\label{Atwomodes}
\end{equation}
The global fermion parity is
\begin{equation}
{\cal P}_0={\rm sign}\,{\rm Pf}\,(-i{\cal A})=-{\rm sign}\,{\rm Det}\,\Gamma.\label{P0globaltwomodes}
\end{equation}

To obtain the average local fermion parity we use that the real $2\times 2$ coupling matrix $\Gamma$ has the singular value decomposition $\Gamma=O_1\,{\rm diag}\,(\kappa_1,\kappa_2)O_2$, with $O_1,O_2$ real orthogonal matrices and $\kappa_1,\kappa_2>0$. The eigenvalues of ${\cal A}$ are $\pm\kappa_1,\pm\kappa_2$. In the flat-band transformation $\{\kappa_1,\kappa_2\}\mapsto\{1,1\}$, which gives
\begin{equation}
\bar{\cal A}=\begin{pmatrix}
0&iO_1O_2\\
-iO_2^{\rm T}O_1^{\rm T}&0
\end{pmatrix}\Rightarrow
[\bar{\cal A}]_{\rm LL}=0\Rightarrow\langle {\cal P}_{\rm L}\rangle=0,\label{barAtwomodesLL}
\end{equation}
so the average fermion parity in a single quantum dot vanishes. This is a manifestation of the Majorana fusion rule \cite{Nay08}: The fusion of the two Majorana zero-modes $\gamma_1$ and $\gamma_2$ produces an \textit{equal-weight} superposition of a state of even and odd fermion parity.\footnote{The converse is not excluded:  $\langle{\cal P}_{\rm L}\rangle= 0$ without an isolated Majorana zero-mode is possible, for example for\\
${\cal A}=i\tiny{\begin{pmatrix}
0&\lambda_1&0&\lambda_2\\
-\lambda_1&0&-\lambda_2&0\\
0&\lambda_2&0&\lambda_1\\
-\lambda_2&0&-\lambda_1&0
\end{pmatrix}}$ with $\lambda_1<\lambda_2$.}

Several recent experimental proposals \cite{Cla17,Aas16,Bee19} are based on the connection between the Majorana fusion rule and vanishing average fermion parity. The implication ``isolated Majorana zero-modes $\Rightarrow$ $\langle{\cal P}_{\rm L}\rangle=0$'' holds if there are only two pairs of Majorana zero-modes. For $N_{\rm L}$ or $N_{\rm R}$ greater than 1 the implication breaks down, as is demonstrated by the following counterexample for $N_{\rm L}=N_{\rm R}=2$: 
\begin{subequations}
\label{testoneway}
\begin{align}
&{\cal A}=\begin{pmatrix}
i\Omega&i\Gamma\\
-i\Gamma^{\rm T}&i\Omega
\end{pmatrix},\;\;\Omega=\begin{pmatrix}
0&0&0&0\\
0&0&0&0\\
0&0&0&1\\
0&0&-1&0
\end{pmatrix},\;\;
\Gamma=\begin{pmatrix}
0&0&1&0\\
0&0&0&1\\
1&0&0&0\\
0&1&0&0
\end{pmatrix},\\
&\Rightarrow\bar{\cal A}=\frac{1}{\sqrt 5}\begin{pmatrix}
i\Omega'&i\Gamma'\\
-i\Gamma'^{\rm T}&i\Omega'
\end{pmatrix},\;\;\Gamma'=2\Gamma,\;\;\Omega'=\begin{pmatrix}
0&-1&0&0\\
1&0&0&0\\
0&0&0&1\\
0&0&-1&0
\end{pmatrix},\\
&\Rightarrow[\bar{\cal A}]_{\rm LL}=\frac{i}{\sqrt 5}\Omega'\Rightarrow{\rm Pf}\,[-i\bar{\cal A}]_{\rm LL}=-\frac{1}{ 5},
\end{align}
\end{subequations}
and hence $\langle{\cal P}_{\rm L}\rangle=-1/ 5$ does not vanish even though each quantum dot has a pair of Majorana zero-modes without intra-dot coupling ($\gamma_1$ and $\gamma_2$ in the left dot, $\gamma_5$ and $\gamma_6$ in the right dot).

Since ${\rm Pf}\,(-i{\cal A})=+1$ the global fermion parity is even, hence the negative sign for $\langle{\cal P}_{\rm L}\rangle$ means that the states with odd-odd occupation numbers in the left and right quantum dot have a greater weight in the ground state than the states with even-even occupation numbers --- even though the fusion of the Majorana modes $\gamma_1$ and $\gamma_2$ would favor equal weight of even and odd fermion parity.

As a check on the formalism, we have also calculated the average fermion parity directly from the many-particle ground state wave function $|\Psi_0\rangle$ of the Hamiltonian $H=\tfrac{1}{2}\gamma\cdot{\cal A}\cdot\gamma$. We find
\begin{align}
|\Psi_0\rangle=\frac{\sqrt{5}}{10}\bigl[2i(a_1^\dagger a_2^\dagger+a_3^\dagger a_4^\dagger)&-(1+\sqrt{5})a_1^\dagger a_3^\dagger\nonumber\\
&-(1-\sqrt{5})a_2^\dagger a_4^\dagger\bigr]|0\rangle,
\end{align}
which indeed gives $\langle{\cal P}_{\rm L}\rangle=-1/ 5$ upon calculation of the expectation value (\ref{PLdef}). 

In this case with $N=N_{\rm L}+N_{\rm R}=4$ electronic levels the size $2^{N-1}=8$ of the basis of many-particle states in the even-parity sector is the same as the size $2N=8$ of the basis of single-particle states, so the two calculations based on Eq.\ (\ref{PLdef}) or on Eq.\ (\ref{generalizedPL}) are equally efficient. For larger $N$ the single-particle approach based on the Pfaffian formula has the more favorable scaling (polynomial instead of exponential).

\section{Random-matrix theory}
\label{sec_RMT}

For a statistical description of the fermion parity fluctuations we apply the methods of random-matrix theory (RMT). In Sec.\ \ref{sec_PLdistribution} we assume a strong mixing of the states in the two quantum dots of Fig.\ \ref{fig_diagram}, and then in Sec.\ \ref{sec_coupling} we consider the opposite regime of weakly coupled quantum dots. We will need results \cite{Dah10} from the RMT in symmetry class DIII, which we summarize in Sec. \ref{sec_SCE}.

\subsection{Skew Circular Real Ensemble}
\label{sec_SCE}

The matrix $[-i\bar{\cal A}]_{\rm LL}$ which in view of Eq.\ (\ref{generalizedPL}) determines the local fermion parity is a $2N_{\rm L}\times 2N_{\rm L}$ submatrix of a matrix ${\cal S}=-i\bar{\cal A}$ that is an antisymmetric (skew-symmetric) element of the real orthogonal group ${\rm O}(2N)$, with $N=N_{\rm L}+N_{\rm R}$. The corresponding ensemble from RMT is the class-DIII circular ensemble, which differs from the class-D circular ensemble by the antisymmetry restriction \cite{Bee15}.  The latter is called the \textit{Circular Real Ensemble} (CRE) and we will refer to the former as the \textit{skew-Circular Real Ensemble} (skew-CRE).\footnote{The qualifier ``real'' for the ${\rm O}(N)$ ensemble is used instead of ``orthogonal'' because the name \textit{Circular Orthogonal Ensemble} (COE) was already used by Dyson \cite{Dys62} for the coset ${\rm U}(N)/{\rm O}(N)$.} The switch from symmetry class D to DIII is remarkable, because class DIII was originally introduced \cite{Alt97} in superconductors with preserved time-reversal symmetry --- which is broken in our physical system.

Two equivalent methods to randomly choose a matrix from the skew-CRE are:
\begin{enumerate}
\item
Generate a real antisymmetric matrix $-i{\cal A}$ with independent Gaussian elements on the upper diagonal (zero mean and unit variance), and perform the flat-band transformation to obtain ${\cal S}=-i\bar{\cal A}$.
\item
Draw a random element $O$ from ${\rm O}(2N)$, uniformly with the invariant Haar measure, and construct
\begin{equation}
{\cal S}=O\begin{pmatrix}
0_{N\times N}&1_{N\times N}\\
-1_{N\times N}&0_{N\times N}
\end{pmatrix}O^{\rm T}.\label{barOconstruction}
\end{equation}
\end{enumerate}
The two methods are equivalent because the distribution $P({\cal A})\propto\exp(\tfrac{1}{4}{\rm Tr}\,{\cal A}^{2})$ as well as the flat-band transformation ${\cal A}\mapsto\bar{\cal A}$ are invariant under orthogonal transformations ${\cal A}\mapsto O{\cal A}O^{\rm T}$, so the matrix $O$ in the decomposition (\ref{barOconstruction}) is distributed according to the invariant Haar measure.

The matrix ${\cal S}$ has the block decomposition
\begin{equation}
{\cal S}=\begin{pmatrix}
{\cal S}_{\rm LL}&{\cal S}_{\rm LR}\\
{\cal S}_{\rm RL}&{\cal S}_{\rm RR}
\end{pmatrix},\;\;{\cal S}_{\rm LL}=[-i\bar{\cal A}]_{\rm LL},\label{calSblock}
\end{equation}
with ${\cal S}_{XY}$ a matrix of dimension $N_{X}\times N_Y$. In the context of scattering problems, where the skew-CRE ensemble was studied previously \cite{Bee15}, this is analogous to a decomposition of the scattering matrix into reflection and transmission matrices. In that context the eigenvalues $\pm i\lambda_n$ of the upper-left submatrix ${\cal S}_{\rm LL}$ correspond to reflection amplitudes.\footnote{Eq.\ (\ref{PlambdaskewCRE}) follows from equation 5 of Ref.\ \citen{Dah10} upon change of variables from transmission probabilities $T_n$ to reflection amplitudes $\lambda_n=\sqrt{1-T_n}$.} Their joint probability distribution in the skew-CRE is known \cite{Dah10},
\begin{align}
&P(\lambda_1,\lambda_2,\ldots \lambda_{N_{\rm min}})\propto\prod_{n}(1-\lambda_n^2)^{|N_{\rm L}-N_{\rm R}|}\prod_{j<k}(\lambda_k^{2}-\lambda_j^{2})^2,\nonumber\\
&\qquad N_{\rm min}=\min(N_{\rm L},N_{\rm R}),\;\;0\leq \lambda_n\leq 1.
\label{PlambdaskewCRE}
\end{align}
If $N_{\rm L}>N_{\rm R}$ there are additionally $2(N_{\rm L}-N_{\rm R})$ trivial eigenvalues pinned at $\pm 1$, not included in the distribution (\ref{PlambdaskewCRE}).

Symmetry class DIII has the $\mathbb{Z}_2$ invariant ${\rm Pf}\,{\cal S}=\pm 1$, which in view of Kitaev's formula (\ref{Kitaevnosign}) is the global fermion parity ${\cal P}_0$. This does not enter in Eq.\ (\ref{PlambdaskewCRE}) because in the skew-CRE the distribution of the $\lambda_n$'s is independent of the $\mathbb{Z}_2$ invariant \cite{Dah10}.

The density $\rho(\lambda)$ of the nontrivial eigenvalues has $\pm\lambda$ symmetry with a three-peak structure: There are two peaks at  the band edges $\pm \lambda_c$, with \cite{Dah10}
\begin{equation}
\lambda_c=(2/N)(N_{\rm L}N_{\rm R})^{1/2},\label{lambdacdef}
\end{equation}
and a peak at the band center\footnote{For the density profile near $\lambda=0$ we can approximate the distribution (\ref{PlambdaskewCRE}) by $P(\{\lambda\})\propto\prod_{j<k}(\lambda_k^{2}-\lambda_j^{2})^2$ and ignore the restriction $|\lambda_n|\leq 1$. The distribution of the $\lambda_n$'s is then identical to the distribution of the energy levels of a Hermitian matrix in symmetry class D, which has the spectral peak (\ref{rholambda}). A Hermitian matrix in class DIII, rather than class D, has a vanishing density of states at zero energy, but this is not relevant for $\rho(\lambda)$.} described by \cite{Alt97,Mehta,Iva02}
\begin{equation}
\rho(\lambda)=\frac{1}{\delta_{\rm eff}}+\frac{\sin(2\pi\lambda/\delta_{\rm eff})}{2\pi \lambda},\;\;\lambda\lesssim 1/\delta_{\rm eff}.\label{rholambda}
\end{equation}
The parameter $\delta_{\rm eff}=\pi/2N_{\rm min}$ is the mean eigenvalue spacing in the center of the band. The peak at $\lambda=0$ is a weak antilocalization effect in the scattering context \cite{Pik12}. 

\begin{figure}[tb]
\centerline{\includegraphics[width=0.9\linewidth]{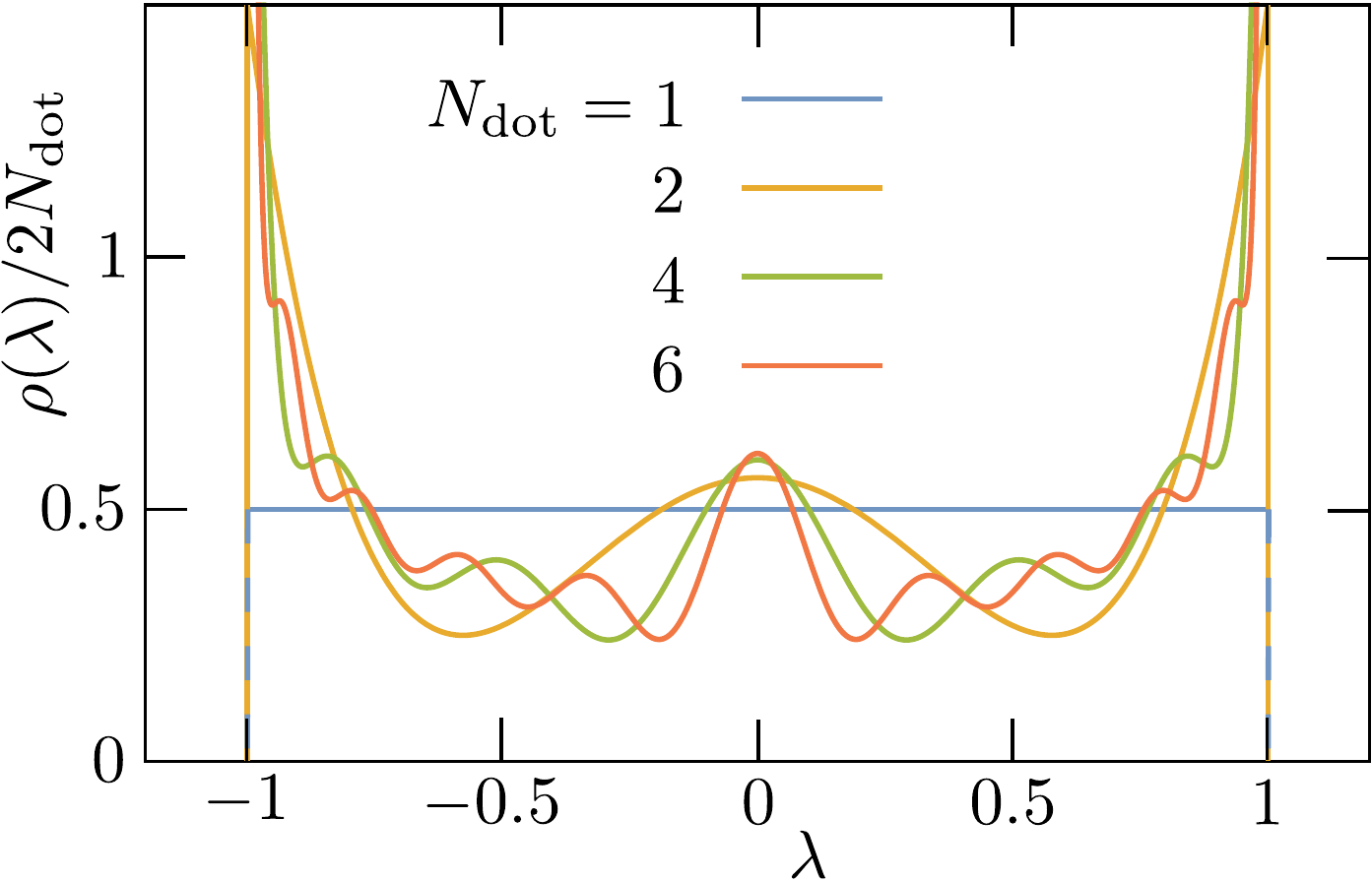}}
\caption{Density $\rho(\lambda)$ of the eigenvalues of the $2N_{\rm L}\times 2N_{\rm L}$ matrix $[-i\bar{\cal A}]_{\rm LL}$ in the skew-CRE, calculated by integration of the distribution (\ref{PlambdaskewCRE}) for $N_{\rm L}=N_{\rm R}=N_{\rm dot}\in\{1,2,4,6\}$. The density has a peak at the band edges and at the band center.
}
\label{fig_rholambda}
\end{figure}

Fig.\ \ref{fig_rholambda} shows the eigenvalue density for $N_{\rm L}=N_{\rm R}=N_{\rm dot}$ ranging from 1 to 6. The three-peaked structure is evident except for $N_{\rm dot}=1$, when the density profile is flat.

\subsection{Distribution of the local fermion parity in the skew-CRE}
\label{sec_PLdistribution}

\begin{figure}[tb]
\centerline{\includegraphics[width=0.9\linewidth]{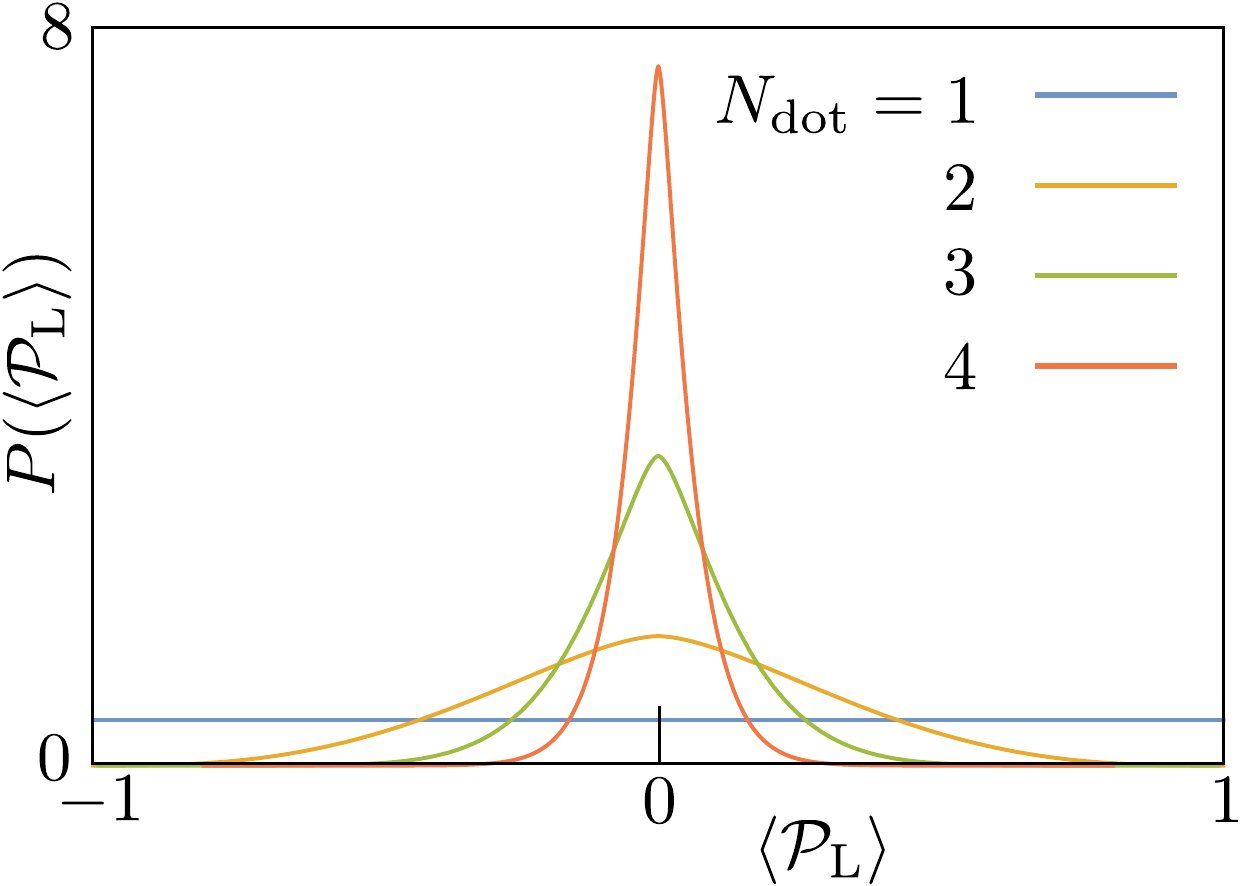}}
\caption{Probability distribution of the local fermion parity in the ensemble of antisymmetric orthogonal matrices (skew-CRE), representative of strongly coupled quantum dots. The curves are calculated from Eq.\ (\ref{PlambdaskewCRE}) for $N_{\rm L}=N_{\rm R}=N_{\rm dot}\in\{1,2,3,4\}$. It takes just a few levels in the quantum dot to have $\langle {\cal P}_{\rm L}\rangle\approx 0$ with high probability, so equal weight of even and odd fermion parity.
}
\label{fig_PPL}
\end{figure}

The peak at $\lambda=0$ in the eigenvalue density $\rho(\lambda)$ increases the probability for vanishing local fermion parity, since
\begin{equation}
|\langle{\cal P}_{\rm L}\rangle|=\prod_{n=1}^{N_{\rm min}}\lambda_n=\sqrt{{\rm Det}\,{\cal S}_{\rm LL}}.\label{PLNmin}
\end{equation}
Indeed, as shown in Fig.\ \ref{fig_PPL}, while the distribution of $\langle{\cal P}_{\rm L}\rangle$ in the skew-CRE is broad for a single electronic level $N_{\rm dot}=1$ in each quantum dot, it quickly narrows to a sharp peak at $\langle{\cal P}_{\rm L}\rangle=0$ with just a few levels --- in accord with numerical calculations reported by Clarke, Sau, and Das Sarma \cite{Cla17}.

The peak at zero $\langle{\cal P}_{\rm L}\rangle\equiv p$ appears as a sharp cusp in Fig.\ \ref{fig_PPL}, it has a logarithmic singularity $\propto (p^2\ln|p|)^{N_{\rm dot}-1}$, for example
\begin{equation}
P(\langle{\cal P}_{\rm L}\rangle=p)=\tfrac{45}{32} \left(1-p^4+4 p^2 \ln|p|\right),\;\; N_{\rm dot}=2,\;\;|p|\leq 1.\label{PpNdot2}
\end{equation}
For large-$N_{\rm dot}$ the width of the distribution becomes exponentially small, as follows from the variance
\begin{equation}
{\rm Var}\,\langle{\cal P}_{\rm L}\rangle=\frac{(2N_{\rm dot})!^3}{(N_{\rm dot})!^2(4N_{\rm dot})!}= \frac{\sqrt{2}}{4^{N_{\rm dot}}}[1+{\cal O}(1/N_{\rm dot})],\label{mu2presult}
\end{equation}
see App.\ \ref{app_determinant}.

We may quantify the effect of the spectral peak in $\rho(\lambda)$ on the distribution of the local fermion parity by comparing with a set of independent $\lambda_n$'s with uniform density. In that uniform case one would have the fermion parity distribution
\begin{equation}
P_{\rm uniform}(\langle{\cal P}_{\rm L}\rangle=p)=\frac{(-\ln|p|)^{N_{\rm dot}-1}}{2(N_{\rm dot}-1)!},\;\;|p|\leq 1,\label{PNdotuniform}
\end{equation}
with a variance $3^{-N_{\rm dot}}$ that decays less rapidly than Eq.\ (\ref{mu2presult}).

\subsection{RMT model of weakly coupled quantum dots}
\label{sec_coupling}

\begin{figure}[tb]
\centerline{\includegraphics[width=0.9\linewidth]{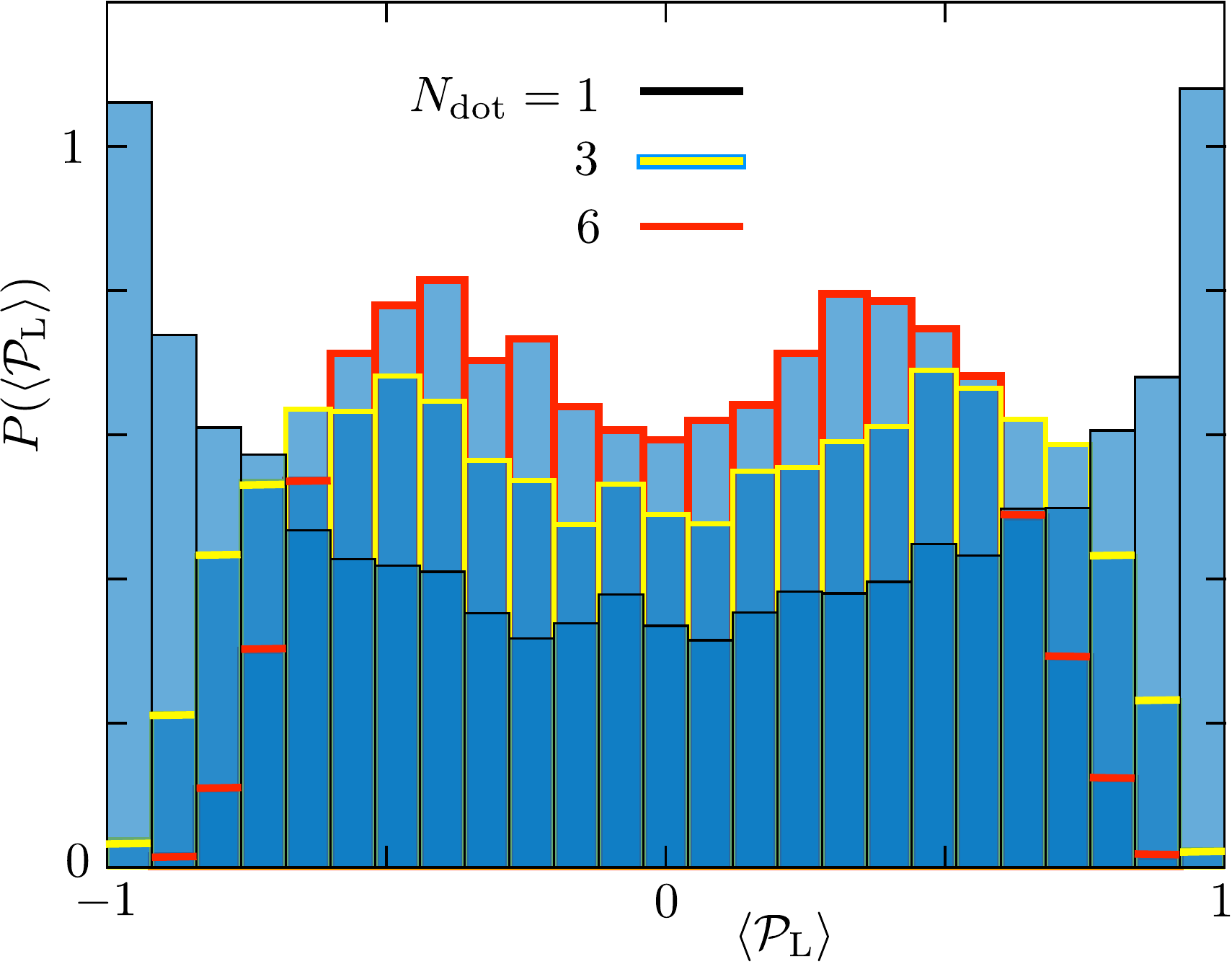}}
\caption{Probability distribution of the local fermion parity for the RMT model (\ref{Aquantumdots}) of two weakly coupled quantum dots, calculated numerically by sampling the Gaussian matrix elements in $\Omega_{\rm L},\Omega_{\rm R},\Gamma$ for $N_{\rm QPC}=1$, $N_{\rm L}=N_{\rm R}=N_{\rm dot}\in\{1,3,6\}$. In contrast to the strongly coupled skew-CRE ensemble of Fig.\ \ref{fig_PPL}, the distribution narrows only slowly with increasing $N_{\rm dot}$.
}
\label{fig_histP}
\end{figure}

\begin{figure}[tb]
\centerline{\includegraphics[width=0.9\linewidth]{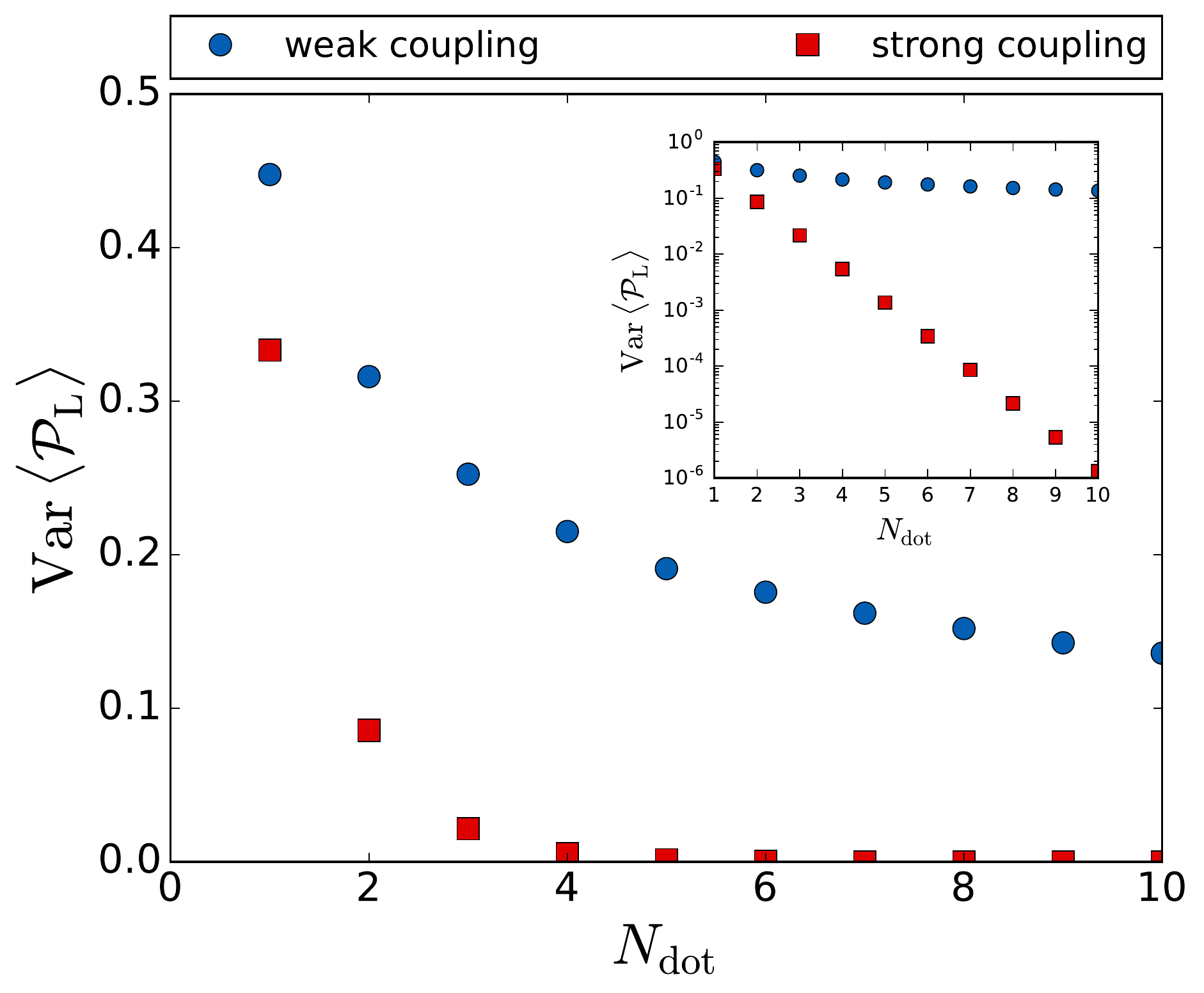}}
\caption{Comparison of the variance of $P(\langle {\cal P}\rangle)$ in the skew-CRE of strongly coupled quantum dots [red data points, calculated from Eq.\ (\ref{mu2presult})] and in the weakly coupled ensemble (blue data points, numerical results for $N_{\rm QPC}=1$). The inset shows that the decay is exponential in both cases, but with widely different decay rates.
}
\label{fig_varP}
\end{figure}

\begin{figure*}[tb]
\centerline{\includegraphics[width=1\linewidth]{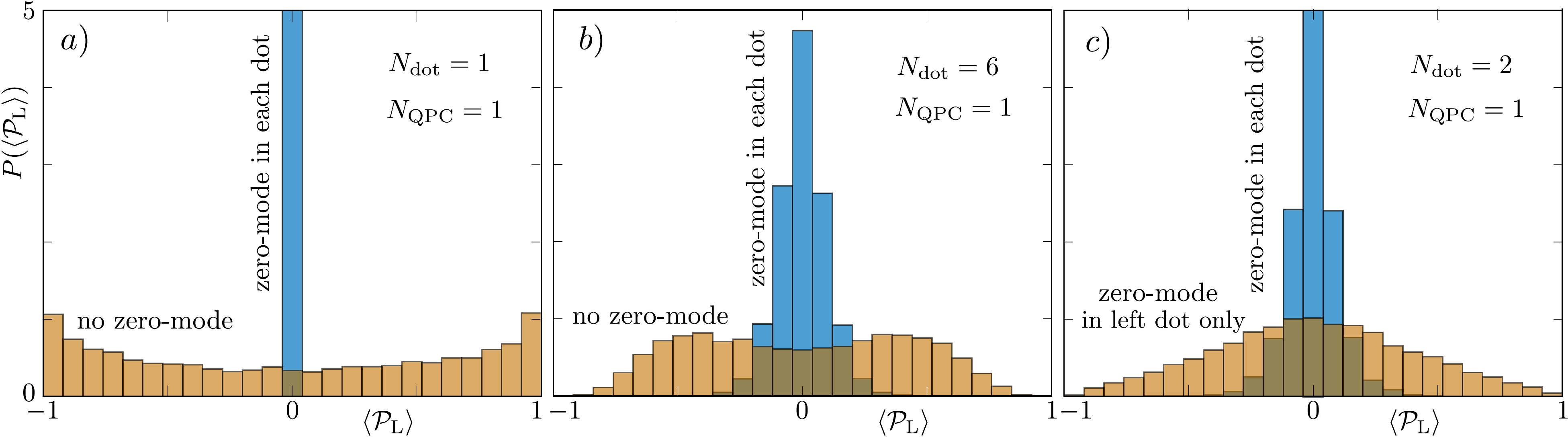}}
\caption{Same as Fig.\ \ref{fig_histP}, but now comparing the situation with or without isolated Majorana zero-modes in a quantum dot. The quantum dots are weakly coupled ($N_{\rm QPC}=1)$ and they have the same number of electronic levels $N_{\rm L}=N_{\rm R}=N_{\rm dot}$. For the blue histograms each quantum dot has a pair of isolated Majorana zero-modes (no intradot coupling, only interdot coupling). For the brown histograms there are either no isolated zero-modes at all (panels a and b), or they are only in one of the two quantum dots (panel c). Weak coupling ensures that the peak at vanishing local fermion parity becomes a distinctive feature of isolated Majorana zero-modes in each quantum dot.
}
\label{fig_comparison}
\end{figure*}

The RMT description in terms of the skew-CRE from the previous subsection assumes a strong (chaotic) mixing in the entire phase space, appropriate for strongly coupled quantum dots. To describe also the weakly coupled regime, we consider an alternative approach where the RMT ensemble is applied to the two quantum dots individually, rather than to the system as a whole.

In the Majorana representation, the Hamiltonian $H=\tfrac{1}{2}\gamma\cdot{\cal A}\cdot\gamma$ of the two coupled quantum dots of Fig.\ \ref{fig_diagram} has the block structure
\begin{equation}
{\cal A}=\begin{pmatrix}
i\Omega_{\rm L}&i\Gamma\\
-i\Gamma^{\rm T}&i\Omega_{\rm R}
\end{pmatrix}.\label{Aquantumdots}
\end{equation}
The real antisymmetric matrices $\Omega_{X}$ of size $2N_{X}\times 2N_{X}$, with $X\in\{\text{R,L}\}$, describe the left and right quantum dot in isolation, while the $2N_{\rm L}\times 2N_{\rm R}$ real matrix $\Gamma$ describes the coupling via a quantum point contact (QPC) with $N_{\rm QPC}$ propagating fermionic modes. In what follows we take $N_{\rm L}=N_{\rm R}=N_{\rm dot}$.

The number $N_{\rm dot}$ counts the number of electronic modes in each quantum dot. One electronic mode $a_n$ corresponds to two Majorana modes $\gamma_{2n-1}$ and $\gamma_{2n}$, according to
\begin{equation}
a_n=(\gamma_{2n-1}+i\gamma_{2n})/\sqrt 2,\label{angammanrelation}
\end{equation}
cf.\ Eq.\ (\ref{gammaUdef}). Because of this double-counting, the mean level spacing $\delta_0$ of eigenstates of $\Omega_X$ is one half the electronic mean level spacing of a quantum dot (taken the same in each dot, for simplicity).

For a statistical description we take independent Gaussian distributions for the two matrices $\Omega_{X}$. Each upper-diagonal matrix element has zero mean and variance  $2N_{\rm dot}\delta_0^2/\pi^2$, corresponding to superconductors in symmetry class D (broken time-reversal and broken spin-rotation symmetry) \cite{Alt97,Bee15}. 

Following Refs.\ \citen{Bar00,Die06}, the quantum dots are coupled by a Gaussian random matrix $\Gamma$ of rank $N_{\rm QPC}$, with elements\footnote{The coupling matrix (\ref{Gammadef}) describes a ballistic point contact. For tunnel coupling, rather than ballistic coupling, the coupling strength $\delta_0/\pi$ is to be multiplied by $T_n^{-1}(2-T_n-2\sqrt{1-T_n})$, with $T_n$ the tunnel probability of the $n$-th mode in the QPC, see Ref.\ \citen{Bee15}.}
\begin{equation}
\Gamma_{kl}=\frac{2N_{\rm dot}\delta}{\pi}\sum_{n=1}^{2N_{\rm QPC}}v^{(n)}_k w^{(n)}_{l},\label{Gammadef}
\end{equation}
in terms of $2N_{\rm QPC}$ real Gaussian vectors $v^{(n)}$ and $w^{(n)}$ of unit average length (each element independently distributed with zero mean and variance $1/2N_{\rm dot}$).

For the weak coupling regime we focus on the case of a single propagating electronic mode in the point contact, $N_{\rm QPC}=1$, corresponding to two propagating Majorana modes. We do not have an analytical solution, so we show numerical results in Fig.\ \ref{fig_histP} for the probability distribution of $\langle{\cal P}_{\rm L}\rangle={\rm Pf}\,(-i\bar{\cal A})$ in the ensemble of random matrices $\Omega_{\rm L}$, $\Omega_{\rm R}$, and $\Gamma$. The variance of the distribution is compared with that in the skew-CRE in Fig.\ \ref{fig_varP}. The two figures show that the distribution of the local fermion parity is much broader when the coupling is via a single-mode point contact.

\section{Effect of an isolated Majorana zero-mode}
\label{sec_isolated}

The random Hamiltonians of the previous section do not contain isolated Majorana zero-modes: the $2N_{\rm dot}$ Majorana modes in each quantum dot have intradot coupling as well as interdot coupling. We may introduce a pair of isolated Majorana zero-modes in a quantum dot by setting to zero one row and one column of the submatrix $\Omega_{\rm L}$ or $\Omega_{\rm R}$ in the Hamiltonian \eqref{Aquantumdots}. (The row and column number should be the same to preserve the antisymmetry of $\Omega_{X}$.) The effect on the distribution of the local fermion parity is shown in Fig.\ \ref{fig_comparison}. The distribution of the local fermion parity is strongly peaked at zero if and only if there is a pair of isolated Majorana zero-modes in each of the two quantum dots.

\section{Conclusion}
\label{sec_discuss}
	
In summary, we have studied the fusion of Majorana zero-modes using a generalization of Kitaev's Pfaffian formula \cite{Kit01} for the global fermion parity of the superconducting ground state, to include local fermion parity fluctuations in an open subsystem. The Pfaffian formula in Eq.\ (\ref{generalizedPL}), and an equivalent formulation from Ref.\ \citen{Bau18}, is computationally efficient since it works with the single-particle (Bogoliubov-De Gennes) Hamiltonian rather than with the many-particle Hamiltonian. One limitation of the single-particle formulation is that it is limited to a mean-field description of the superconductor --- in particular we are assuming that the quantum dots in the geometry of Fig.\ \ref{fig_diagram} have a sufficiently large capacitance that Coulomb charging energies can be neglected.

The Pfaffian fermion parity formula is particularly suited to an analysis in terms of random-matrix theory, in an ensemble of antisymmetric matrices \cite{Bee15}. For strongly coupled quantum dots the circular ensemble in symmetry class DIII is the appropriate ensemble, which allows for analytical results for the statistical distribution of the local fermion parity.  There is no eigenvalue repulsion at the particle-hole symmetry point in such an ensemble \cite{Alt97}, and the resulting accumulation of near-zero eigenvalues enforces a nearly equal-weight superposition of even and odd fermion parity in a quantum dot.

This is a nontopological mechanism for vanishing expectation value $\langle{\cal P}_{\rm L}\rangle\approx 0$ of the local fermion parity. The Majorana fusion rule provides a fundamentally different, topological mechanism \cite{Nay08}: The merging or ``fusion'' of two isolated Majorana zero-modes (``isolated'' in the sense of zero intradot coupling, while allowing for interdot coupling) also favors a vanishing $\langle{\cal P}_{\rm L}\rangle$ because the two fusion channels, with or without an unpaired quasiparticle, have equal weight.

To carry out such a fusion experiment it is proposed \cite{Aas16} that one would rapidly decouple the subsystems, on a time scale $\tau_c$ sufficiently short that quasiparticles from the environment cannot leak in. The complication \cite{Cla17} is that even if there are isolated Majorana zero-modes, the presence of even a small number $N_{\rm dot}$ of higher levels at energies below $\hbar/\tau_c$ may hide the presence of the zero-modes by favoring $\langle{\cal P}_{\rm L}\rangle\approx 0$ (see Fig.\ \ref{fig_PPL}).

Fig.\ \ref{fig_comparison} illustrates our proposal to distinguish the two mechanisms for vanishing local fermion parity: A low-rank coupling between the quantum dots, via a single-mode quantum point contact, suppresses the nontopological effect from levels at nonzero energy, without affecting the topological effect from the fusion of isolated Majorana zero-modes.

\begin{acknowledgements}
We have benefited from discussions with A. R. Akhmerov and Y. Herasymenko. J. H. H. Perk alerted us to the connection between Eqs.\ (\ref{generalizedPL}) and (\ref{Wick}). This research was supported by the Netherlands Organization for Scientific Research (NWO/OCW)  and by the European Research Council.
\end{acknowledgements}

\appendix

\section{Derivation of the Pfaffian formula from Klich's counting statistics theory}
\label{App_Klich}

We follow the steps of Klich's theory of counting statistics of paired fermions \cite{Kli14}, to reproduce his result for $\langle{\cal P}_{\rm L}\rangle^2$. Then we will resolve the sign ambiguity to arrive at Eq.\ (\ref{generalizedPL}) for $\langle{\cal P}_{\rm L}\rangle$. An equivalent formula is obtained by a different method in Ref.\ \citen{Bau18}, Appendix B.

The superconductor in Fig.\ \ref{fig_diagram} is assumed to be an isolated system, so that the global fermion parity does not fluctuate. For the derivation of the subsystem fermion parity formula (\ref{generalizedPL}) it is convenient to start from the more general case that the superconductor is in contact with a reservoir in thermal equilibrium at temperature $T$. We will then take the $T\rightarrow 0$ limit at the end of the calculation in order to describe an isolated system.

At inverse temperature $\beta=1/k_{{\text B}}T$ the average fermion parity $\langle{\cal P}_{\text L}\rangle$ of subsystem L (the left quantum dot in Fig.\ \ref{fig_diagram}) is given by the trace of the equilibrium density matrix 
\begin{equation}
\rho_{\text{eq}}=\frac{1}{Z} e^{-\beta H},\;\;Z=\text{Tr}\,\rho_{\text{eq}},\label{rhoeqdef}
\end{equation}
acting on the fermion parity operator
\begin{equation}
{\cal P}_{\text L}=\exp\left({i\pi\sum_{n\in\text{L}} a^\dagger_na^{\vphantom{\dagger}}_n}\right).\label{PLgammadef}
\end{equation}
Because $H=\tfrac{1}{2}\sum_{n,m}{\cal A}_{nm}\gamma_n\gamma_m$ in the basis of Majorana operators $\gamma_n$, and $a^\dagger_na^{\vphantom{\dagger}}_n=i\gamma_{2n-1}\gamma_{2n}+\tfrac{1}{2}$, this can be written as
\begin{align}
\langle{\cal P}_{\text L}\rangle={}&\frac{e^{i\pi N_{\text{L}}/2}}{Z}{\rm Tr}\,\left[\exp\left(-\tfrac{1}{2}\beta \sum_{n,m}{\cal A}_{nm}\gamma_n\gamma_m\right)\right.\nonumber\\
&\times\left.\exp\left(-\tfrac{1}{2}i\pi\sum_{n,m}(\sigma_y\otimes P_{\text{L}})_{nm}\gamma_n\gamma_m\right)\right].
\end{align}
The matrix $\sigma_y$ is a Pauli matrix and the operator $P_{\text L}$ projects onto $N_{\text L}$ fermionic modes in subsystem L.

Application of the identity \cite{Kli14} 
\begin{equation}
\left[{\rm Tr}\,\prod_{k}e^{\gamma\cdot O_k\cdot\gamma}\right]^2={}e^{\sum_{k}{\rm Tr}\,O_k}\,{\rm Det}\,\left(1+\prod_{k}e^{O_k-O_k^{\rm T}}\right),\label{Klich2}
\end{equation}
results in
\begin{align}
\langle{\cal P}_{\text L}\rangle^2&=e^{i\pi N_{\text{L}}}\,\frac{{\rm Det}\,\left[1+\exp\left(-\beta  {\cal A}\right)\exp\left(-i\pi\sigma_y\otimes P_{\text{L}}\right)\right]}{{\rm Det}\,\left[1+\exp\left(-\beta  {\cal A}\right)\right]}\nonumber\\
&=(-1)^{N_{\text{L}}}\,{\rm Det}\,\left[1-\frac{2}{1+\exp\left(\beta  {\cal A}\right)} (\sigma_0\otimes P_{\text{L}})\right].\label{calPXresult}
\end{align}
In the second equality we made use of the identity
\begin{equation}
e^{i\chi\,\sigma_y\otimes P_{\text{L}}}=1+\sigma_0\otimes P_{\text{L}}(\cos\chi-1)+i\sigma_y\otimes P_{\text{L}}\sin\chi,\label{identity}
\end{equation}
with $\chi=\pi$. (The matrix $\sigma_0=\sigma_y^2$ is the $2\times 2$ unit matrix.) Note that, in a basis of energy eigenstates of the BdG Hamiltonian, the operator $(1+e^{\beta  {\cal A}})^{-1}$ is the Fermi function $f(E)=(1+e^{\beta E})^{-1}$. 

Eq.\ (\ref{calPXresult}) is Klich's result for the square of the average fermion parity (equation 84 in Ref.\ \citen{Kli14}). Klich shows how the sign of $\langle {\cal P}_{\rm L}\rangle$ can be recovered if the determinant is known analytically as a function of the matrix elements. Here we take a different route, more suitable for numerical calculations, which gives the sign directly upon evaluation of a Pfaffian instead of a determinant.

Any $2N\times 2N$ imaginary anti-symmetric matrix ${\cal A}$ can be decomposed as
\begin{equation}
{\cal A}=iO(J\otimes {\cal E})O^{\text{T}},\;\;J=\begin{pmatrix}
0&1\\
-1&0
\end{pmatrix},\label{calAdecomp}
\end{equation}
where $O$ is a $2N\times 2N$ real orthogonal matrix and\\ ${\cal E}=\text{diag}\,(E_1,E_2,\ldots E_N)$ is an $N\times N$ real diagonal matrix. Substitution into Eq.\ (\ref{calPXresult}) gives
\begin{align}
&\langle{\cal P}_{\text L}\rangle^2=(-1)^{N_{\text{L}}}\,{\rm Det}\,\left[1-O\frac{2}{1+\exp\left(i\beta J\otimes{\cal E}\right)} O^{\text{T}}(\sigma_0\otimes P_{\text{L}})\right]\nonumber\\
&\quad=(-1)^{N_{\text{L}}}\,{\rm Det}\,\left[1-O[1-iJ\otimes\tanh(\tfrac{1}{2}\beta{\cal E})] O^{\text{T}}(\sigma_0\otimes P_{\text{L}})\right].
\end{align}

This may be written in a more compact form by defining the restriction $[M]_{\text{LL}}$ of a $2N\times 2N$ matrix $M$ to the $2N_{\text{L}}\times 2N_{\text{L}}$ submatrix of modes in region L,
\begin{align}
\langle{\cal P}_{\text L}\rangle^2&=(-1)^{N_{\text{L}}}\,{\rm Det}\,\bigl[O[iJ\otimes\tanh(\tfrac{1}{2}\beta{\cal E})]O^{\text{T}}\bigr]_{\text{LL}}\nonumber\\
&={\text{Det}}\,\bigl[O[J\otimes\tanh(\tfrac{1}{2}\beta{\cal E})]O^{\text{T}}\bigr]_{\text{LL}}.
\label{calPXresultprojectedMajorana}
\end{align}
Note that, because of the submatrix restriction, the product rule ${\text{Det}}\,(AB)=({\text{Det}}\,A)({\text{Det}}\, B)$ cannot be applied to ${\text{Det}}[AB]_{\text{LL}}$, so the orthogonal matrix $O$ cannot be cancelled with the inverse $O^{\text{T}}$.

We have now arrived at the determinant of a real antisymmetric matrix, hence we can take the square root without introducing branch cuts,
\begin{equation}
\langle{\cal P}_{\text L}\rangle={\text{Pf}}\,\bigl[O[J\otimes\tanh(\tfrac{1}{2}\beta{\cal E})]O^{\text{T}}\bigr]_{\text{LL}}.\label{PfaffianfiniteT}
\end{equation}
In the zero-temperature, $\beta\rightarrow\infty$ limit this reduces to
\begin{equation}
\langle{\cal P}_{\text L}\rangle={\text{Pf}}\,\bigl[O[J\otimes(\,\text{sign}\,{\cal E})]O^{\text{T}}\bigr]_{\text{LL}},\label{PfaffianzeroT}
\end{equation}
which is Eq.\ (\ref{generalizedPL}) with $-i\bar{\cal A}=O[J\otimes(\text{sign}\,{\cal E})]O^{\text{T}}$. Kitaev's formula (\ref{P0Adef}) for the global ground-state fermion parity is recovered when L is the entire isolated system. This correspondence also identifies $\sqrt\text{Det}$ with $+\text{Pf}$ rather than with $-\text{Pf}$.

\section{Moments of determinants of antisymmetric random matrices}
\label{app_determinant}

In Sec.\ \ref{sec_PLdistribution} we used a formula for the average determinant of a submatrix (a principal minor) of an antisymmetric real orthogonal matrix. This would seem like a classic result in RMT, but we have not found it in the literature on such matrices \cite{Girko,For09,Ede15}. We therefore give the derivation in this appendix, and for completeness and reference also derive the corresponding result for antisymmetric Hermitian matrices.

\subsection{Principal minor of antisymmetric orthogonal matrix}

Consider a $2N\times 2N$ antisymmetric real orthogonal matrix ${\cal S}$, with a uniform distribution in ${\rm O}(2N)$ subject to the antisymmetry constraint. This is the class-DIII circular ensemble of RMT \cite{Alt97,Bee15}, referred to as the skew-Circular Real Ensemble (skew-CRE) in the main text.\footnote{The antisymmetric orthogonal matrices form a disconnected set in ${\rm O}(2N)$, distinguished by the sign of the Pfaffian. For the probability distribution (\ref{eq:jpdf_SCRE}) it does not matter whether or not we restrict the ensemble to ${\rm Pf}\,{\cal S}=\pm 1$.}

The $2N_{\rm L} \times 2N_{\rm L}$ upper-left submatrix ${\cal S}_{\rm LL}$ has eigenvalues $\pm i \lambda_n$, $0\leq \lambda_n\leq 1$. Denoting $N_{\rm R} = N-N_{\rm L}$ and $N_{\rm min}=\min(N_{\rm L},N_{\rm R})$, we have that $N-N_{\rm min}$ of the $\lambda_n$'s are pinned to $+1$. The set $\{\lambda_n\}=\{\lambda_1,\lambda_2,\ldots\lambda_{N_{\rm min}}\}$ can vary freely in the interval $[0,1]$, with joint probability distribution \cite{Dah10}
\begin{equation}
  \label{eq:jpdf_SCRE}
  P(\{ \lambda_n \}) \propto
  \prod_{n} (1-\lambda_n^2)^{|{N_{\rm L} - N_{\rm R}}|}\prod_{i<j} (\lambda_i^2 - \lambda_j^2)^2
  .
\end{equation}
The determinant of ${\cal S}_{\rm LL}$ is a principal minor given by
\begin{equation}
  {\rm Det}\, {\cal S}_{\rm LL}
  = \prod_{n=1}^{N_{\rm L}} (i \lambda_n) (-i \lambda_n)
  = \prod_{n=1}^{N_{\rm min}} \lambda_n^2
  .
\end{equation}
We seek the moments $\mu_q=\mathbb{E}\bigl[({\rm Det}\, {\cal S}_{\rm LL})^q\bigr]$ of this determinant in the skew-CRE.

For that purpose we make a change of variables from $\lambda_n$ to $  R_n = \lambda_n^2\in[0,1]$,
with distribution
\begin{equation}
  P(\{ R_n \}) \propto
  \prod_{n} R_n^{-1/2} (1-R_n)^{|{N_{\rm L}-N_{\rm R}}|}\prod_{i<j} (R_i-R_j)^2
  .
\end{equation}
We can then compute the moments
of the determinant from
\begin{equation}
 \mu_q
  = \frac{
    \displaystyle
   \int_0^1 d\{R_n\}
    \prod_{i<j} (R_i-R_j)^2
    \prod_n (1-R_n)^{|{N_L-N_R}|} R_n^{q-1/2}
  }
  {
    \displaystyle
    \int_0^1 d \{R_n\}
    \prod_{i<j} (R_i-R_j)^2
    \prod_n (1-R_n)^{|{N_L-N_R}|} R_n^{-1/2}
  }
  ,
\end{equation}
where we abbreviated $\int_0^1 d\{R_n\}= \int_0^1 d R_1 \cdots \int_0^1 d R_{N_{\rm min}}$.

These socalled Selberg integrals have a closed-form expression \cite{Forrester},
\begin{equation}
\label{eq:MomSkCRE}
    \mu_q=
    \prod_{j=0}^{N_{\rm min}-1}
    \frac{
      \Gamma\left( \max(N_{\rm L},N_{\rm R})+j+\frac{1}{2} \right)
      \Gamma\left( q+j+\frac{1}{2} \right)
    }
    {
      \Gamma\left( \max(N_{\rm L},N_{\rm R})+q+j+\frac{1}{2} \right)
      \Gamma\left( j+\frac{1}{2} \right)
    }
    .
\end{equation}
For the first few moments, Eq.~(\ref{eq:MomSkCRE}) reduces to
\begin{align}
 &\mu_1= 
  \frac{(2 N_{\rm L})! (2N_{\rm R})! N!}{N_{\rm L}! N_{\rm R}! (2N)!}
  ,\label{mu1result}\\
&  \mu_2
  = \frac{(2N_{\rm L}+1)(2N_{\rm R}+1)}{2N+1}  \mu_1^2
  .
\end{align}
Eq.\ (\ref{mu2presult}) in the main text is Eq.\ (\ref{mu1result}) for $N_{\rm L}=N_{\rm R}=N_{\rm dot}=N/2$.

\subsection{Antisymmetric Hermitian matrix}

A similar calculation can be carried out for moments of the determinant of a $2N\times 2N$ antisymmetric Hermitian matrix ${\cal A}$, in the Gaussian ensemble of independent upper-diagonal elements with a normal distribution (zero mean and unit variance). 

The $2N$ eigenvalues come in pairs $\pm\lambda_n$. The $N$ eigenvalues $\lambda_n\geq 0$ have the joint distribution \cite{Mehta}
\begin{equation}
  \label{eq:DistrEig}
  P(\{ \lambda_n \}) \propto
    \prod_{n}  e^{- \lambda_n^2/2}\prod_{i<j} (\lambda_i^2 - \lambda_j^2)^2
.
\end{equation}
The determinant is
\begin{equation}
  {\rm Det}\,{\cal A} =(-1)^N \prod_{n=1}^N \lambda_n^2
  .
\end{equation}

Let us introduce the variables $x_n =\lambda_n^2/2\geq 0$, with distribution
\begin{equation}
  P(\{ x_n \}) \propto
   \prod_n
  x_n^{-1/2} e^{- x_n}\prod_{i<j} (x_i - x_j)^2.
\end{equation}
The $q$-th moment $\mu_q$ of the determinant of ${\cal A}$ is given by
\begin{equation}
  \mu_q= (-2)^{Nq}
  \frac{\displaystyle
    \int_0^\infty d\{x_n\} 
    \prod_{i<j} (x_i-x_j)^2 \prod_{n} x_n^{q - 1/2} e^{-x_n}
  }{\displaystyle
    \int_0^\infty d\{x_n\}
    \prod_{i<j} (x_i-x_j)^2 \prod_{n} x_n^{- 1/2} e^{-x_n}
  }
  ,
\end{equation}
with $\int_0^\infty d\{x_n\}=\int_0^\infty dx_1\cdots\int_0^\infty dx_N$.
This is the ratio of normalisation constants of Laguerre
distributions, which is known \cite{Forrester}.
We thus obtain
\begin{equation}
  \label{eq:ResMomDetA}
    \mu_q
    = (-2)^{Nq}
    \prod_{j=0}^{N-1}
    \frac{\Gamma\left(q+N-j-\frac{1}{2}\right)}
    {\Gamma\left(N-j-\frac{1}{2}\right)}
    .
\end{equation}

For $q=1,2$ this reduces to
\begin{equation}
\begin{split}
&  \label{eq:MoyDetA}
  \mu_1= (-1)^N
  \frac{(2N)!}{2^N N!},\;\;\mu_2= 
  \frac{(2N+1)!(2N)!}{2^{2N} (N!)^2},\\
  &\Rightarrow{\rm Var}\,({\rm Det}\, {\cal A}) = 2 N [\mathbb{E}({\rm Det}\,{\cal A})]^2.
  \end{split}
  \end{equation}
The average determinant of antisymmetric Hermitian matrices increases exponentially with $N$,
\begin{equation}
\mu_1=\sqrt{2}(-2/e)^N  N^N[1+{\cal O}(1/N)],
\end{equation}
in contrast to the exponential decay for antisymmetric orthogonal matrices, cf.\ Eq.\ (\ref{mu2presult}).


\begin{thebibliography}{99}
\bibitem{Bal06} A. V. Balatsky, I. Vekhter, and Jian-Xin Zhu, \textit{Impurity-induced states in conventional and unconventional superconductors}, Rev. Mod. Phys. \textbf{78}, 373 (2006).
\bibitem{Sak70} A. Sakurai, \textit{Comments on superconductors with magnetic impurities}, Prog. Theor. Phys. \textbf{44}, 1472 (1970).
\bibitem{Kit01} A. Yu. Kitaev, \textit{Unpaired Majorana fermions in quantum wires}, Phys. Usp. \textbf{44} (suppl.), 131 (2001).
\bibitem{Alt97} A. Altland and M. R. Zirnbauer, \textit{Nonstandard symmetry classes in mesoscopic normal-superconducting hybrid structures}, Phys. Rev. B \textbf{55}, 1142 (1997).
\bibitem{Bee15} C. W. J. Beenakker, \textit{Random-matrix theory of Majorana fermions and topological superconductors}, Rev. Mod. Phys. \textbf{87}, 1037 (2015).
\bibitem{Cla17} D. J. Clarke, J. D. Sau, and S. Das Sarma, \textit{Probability and braiding statistics in Majorana nanowires}, Phys. Rev. B \textbf{95}, 155451 (2017).
\bibitem{Aas16} D. Aasen, M. Hell, R. V. Mishmash, A. Higginbotham, J. Danon, M. Leijnse, T. S. Jespersen, J. A. Folk, C. M. Marcus, K. Flensberg, and J. Alicea, \textit{Milestones toward Majorana-based quantum computing}, Phys. Rev. X \textbf{6}, 031016 (2016).
\bibitem{Per77} J. H. H. Perk and H. W. Capel, \textit{Time-dependent $xx$-correlation functions in the one-dimensional $xy$-model}, Physica \textbf{89A}, 264 (1977).
\bibitem{Per84} J. H. H. Perk, H. W. Capel, G. R. W. Quispel, and F. W. Nijhoff, \textit{Finite-temperature correlations for the Ising chain in a transverse field}, Physica \textbf{123A}, 1 (1984).
\bibitem{Bra05} S. Bravyi, \textit{Lagrangian representation for fermionic linear optics}, Quantum Inf. Comp. \textbf{5}, 216 (2005).
\bibitem{Bau18} B. Bauer, T. Karzig, R. V. Mishmash, A. E. Antipov, and J. Alicea, \textit{Dynamics of Majorana-based qubits operated with an array of tunable gates}, SciPost Phys. \textbf{5}, 004 (2018).
\bibitem{Kli14} I. Klich, \textit{A note on the full counting statistics of paired fermions}, J. Stat. Mech. P11006 (2014). When comparing formulas, note that Klich has a factor of two in the anticommutator of Majorana operators.
\bibitem{Bud13} J.-C. Budich and E. Ardonne, \textit{Equivalent topological invariants for one-dimensional Majorana wires in symmetry class D}, Phys. Rev. B \textbf{88}, 075419 (2013).
\bibitem{DeG66} P. G. De Gennes, \textit{Superconductivity of Metals and Alloys} (Benjamin, New York, 1966).
\bibitem{Rub11} J. Rubow and U. Wolff, \textit{A factorization algorithm to compute Pfaffians}, Comp. Phys. Comm. \textbf{182}, 2530 (2011).
\bibitem{Wim12} M. Wimmer, \textit{Efficient numerical computation of the Pfaffian for dense and banded skew-symmetric matrices}, ACM Trans. Math. Software \textbf{38}, 30 (2012).
\bibitem{Nay08} C. Nayak, S. H. Simon, A. Stern, M. Freedman, and S. Das Sarma, \textit{Non-Abelian anyons and topological quantum computation}, Rev. Mod. Phys. \textbf{80}, 1083 (2008).
\bibitem{Bee19} C. W. J. Beenakker, A. Grabsch, and Y. Herasymenko, \textit{Electrical detection of the Majorana fusion rule for chiral edge vortices in a topological superconductor}, SciPost Phys. \textbf{6}, 022 (2019).
\bibitem{Dah10} J. P. Dahlhaus, B. B\'{e}ri, and C. W. J. Beenakker, \textit{Random-matrix theory of thermal conduction in superconducting quantum dots}, Phys. Rev. B \textbf{82}, 014536 (2010).
\bibitem{Dys62} F. J. Dyson, \textit{The threefold way: Algebraic structure of symmetry groups and ensembles in quantum mechanics}, J. Math. Phys. \textbf{3}, 1199 (1962).
\bibitem{Mehta} M. L. Mehta, \textit{Random Matrices} (Elsevier, Amsterdam, 2004).
\bibitem{Iva02} D. A. Ivanov, \textit{The supersymmetric technique for random-matrix ensembles with zero eigenvalues}, J. Math. Phys. \textbf{43}, 126 (2002). 
\bibitem{Pik12} D. I. Pikulin, J. P. Dahlhaus, M. Wimmer, H. Schomerus, and C. W. J. Beenakker, \textit{A zero-voltage conductance peak from weak antilocalization in a Majorana nanowire}, New J. Phys.\textbf{14}, 125011 (2012).
\bibitem{Bar00} C. I. Barbosa, T. Guhr, and H. L. Harney, \textit{Impact of isospin breaking on the distribution of transition probabilities}, Phys. Rev. E \textbf{62}, 1936 (2000).
\bibitem{Die06} B. Dietz, T. Guhr, H. L. Harney, and A. Richter, \textit{Strength distributions and symmetry breaking in coupled microwave billiards}, Phys. Rev. Lett. \textbf{96}, 254101 (2006).
\bibitem{Forrester} P. J. Forrester, \textit{Log-Gases and Random Matrices} (Princeton, 2010).
\bibitem{Girko} V. L. Girko, \textit{Theory of Random Determinants} (Springer, Berlin, 1990). 
\bibitem{For09} P. J. Forrester and E. Nordenstam, \textit{The anti-symmetric GUE minor process}, Mosc. Math. J. \textbf{9}, 749 (2009).
\bibitem{Ede15} A. Edelman and M. La Croix, \textit{The singular values of the GUE (Less is More)}, Random Matrices: Theory and Applications \textbf{04}, 1550021 (2015).
\end{thebibliography}
\end{document}